\documentclass[twocolumn,prb,showpacs,letterpaper,amsfonts]{revtex4-1}
\usepackage[utf8]{inputenc}
\usepackage{dcolumn}% Align table columns on decimal point
\usepackage{bm}% bold math
\usepackage{pstricks}
\usepackage{amsmath,amssymb,amsbsy}
\usepackage{graphicx,subfigure}
\usepackage{hyperref}
\usepackage{multirow}

\begin{document}

\title{Stability of boron nitride bilayers: Ground state energies, interlayer distances, and
tight-binding description}


\author{R. M. Ribeiro and N. M. R. Peres}

\affiliation{Department of Physics and Center of Physics,
University of Minho, PT-4710-057, Braga, Portugal}

\date{\today}

%-----------------------------------------------------------
% Abstract
%-----------------------------------------------------------
\begin{abstract}
We have studied boron nitride monolayer and bilayer band structures. For
bilayers, the ground state energies of the different five stackings are computed using DFT
in order to determine the most stable configuration. Also, the interlayer distance
for the five different types of stacking in which boron-nitride bilayers can
be found is determined. Using a minimal tight binding model for the band
structures of boron nitride bilayers, the hopping parameters
and the onsite energies have been extracted by fitting a tight binding empirical model to the
DFT results.
\end{abstract}

\maketitle
%-----------------------------------------------------------
% Main matter
%-----------------------------------------------------------

%------------------------------------------------------------------------------------------
%{\bf Introduction.}
\section{Introduction}
%------------------------------------------------------------------------------------------
It was shown, since the early days of graphene research, that other layered materials
can be exfoliated by micromechanical cleavage, following exactly the same
procedure used to isolate graphene from graphite, giving birth, in this way,
to new two-dimensional crystals, which will trigger, with certainty, new
scientific investigations. \cite{Novoselov2005}
Among these  new materials there are several dichalcogenides, complex oxides, and
boron nitride. It was also shown that these new two-dimensional forms of condensed
matter show high crystal quality and macroscopic
continuity. \cite{Novoselov2005}

This new world of two-dimensional crystals is still much unexplored,
both in what concerns the fundamental properties of these systems and
their potential applications.
Chief among these two-dimensional crystals, and excluding
graphene for obvious reasons, boron nitride
is emerging as a system which is finding applications as a scaffold for graphene
in electronic devices. The fact that boron nitride is an insulator, having a large
energy gap ($\sim 5-7$ eV),
together with its high degree of purity makes it a perfect membrane to isolate
graphene from the imperfect gate dielectrics used so far.
Since both systems, that is, graphene and boron nitride,
interact weakly,  two-dimensional crystals of the latter
can be used as a buffer layer between a substrate and graphene,
leaving its intrinsic electronic properties unafected by disorder
associated with the dielectric surface. As a consequence, large electronic
mobilities, of the same order of magnitude of those found in suspended graphene samples,
have been measured in graphene on top of boron nitride.
\cite{Dean2010}
Graphene devices on hexagonal BN ($h$-BN) have been shown to exhibit enhanced mobility,
reduced carrier inhomogeneity, and reduced intrinsic doping,
 in comparison with graphene laying directly on top SiO$_2$.\cite{Dean2010}
Additionally, this new experimental setup allowed
the measurement of the fractional quantum Hall effect using the four probes
geometry, exposing a buried panoply of fractional filling factors,
 still waiting for deep
fundamental justification.
Other applications of BN can be envisioned, ranging
from spacers to tunneling barriers.

Since these first applications of BN in fundamental research that
the interest in single layer boron nitride (sBN) has been increasing steadily.
In the recent past several experimental studies have been performed,
\cite{Dean2010,Han2008, Meyer2009, Warner2010, Lee2010, Gorbachev2010,Song2010}
including TEM\cite{Han2008,Meyer2009, Warner2010} and AFM\cite{Lee2010} characterization
of the material's surface, optical and Raman\cite{Gorbachev2010} spectroscopy, and
intentional damage sBN's surface.\cite{Meyer2009,Song2010}
The main experimental
approaches for producing sBN have been both mechanical\cite{Novoselov2005,Meyer2009,Dean2010,Lee2010} and chemical\cite{Warner2010} exfoliation. Others chemical methods are also available now.\cite{Han2008,Song2010}
In addition to the discovery that sBN is an excellent substrate for graphene electronic devices, \cite{Dean2010} the fabrication of large area (several cm$^2$) BN layers,\cite{Song2010} opens
the possibility of mass production of this new two dimensional material, at the time of writing quite an expensive one.

In parallel to the experimental
investigations, some theoretical studies on sBN and bilayer boron nitride (bBN) have been
performed;
Topsakal \textit{et al.}\cite{Topsakal2008} made a density functional theory (DFT) study of
the electronic, magnetic, and elastic properties of sBN and boron nitride nanoribbons.
Giovannetti \textit{et al.}\cite{Giovannetti2007} studied graphene on the top of $h$-BN by DFT.
These authors concluded that graphene opens a small gap
for three orientations of graphene on the top of $h$-BN, and that the most stable orientation was an AB stacking with the carbon atoms on the top of the boron atoms.
Following the investigations by Giovannetti \textit{et al.},
S\l{}awi\'{n}ska \textit{et al.}\cite{Slawinska2010} studied
graphene AB stacking on a sBN, using both tight-binding (TB) and DFT.
They confirmed the opening of the gap and fitted the TB parameters to the DFT calculations.
The difficulty with the
aforementioned calculations is the fact that they assume graphene and sBN unit cells
of the same size and perfectly oriented. We note in passing that the opening
of a gap in graphene's spectrum, when this material is laying on top of sBN,
has been speculated to exist in the past and the transport properties
of graphene under these conditions have been computed. \cite{zecarlos}

Despite
the above theoretical studies, it is however experimentally known
that graphene on the top of  sBN
has in general a random crystallographic
orientation, that is, there is no preferential orientation
of graphene's lattice relatively to that of sBN. \cite{Dean2010}
In addition, no energy gap in graphene's spectrum has been measured so far. \cite{Dean2010}

Marom \textit{et al.}\cite{Marom2010} studied the
interlayer sliding of one layer of BN on the top of another layer of BN
using DFT with van der Walls corrections.
They found that the lower energy configurations (AA$'$ and AB) of bBN differ in energy by an
amount smaller than the accuracy limits of their calculations.

In this paper we investigate the structural nature of the ground state of boron nitride bilayers
using density functional theory.
We consider the different possible stacking of the individual
BN planes, compute their energy ground states and band structures,
and fit the TB
hopping parameters and on-site energies to the DFT results.
Graphene bilayer can have two types of stacking: AA and AB (Bernal stacking),
with the AB structure being the most stable one.
BN bilayers have richer structural possibilities;
given that there are two types of atoms, there are five possible bilayer stacking:
two AA and three AB.
Experiments show that both types of stacking exist when one considers  multilayer BN.\cite{Warner2010}
It is then important to clarify what can be expected in what concerns the energetic
stability of the different stacking
when one takes a bilayer as an isolated
system.

%------------------------------------------------------------------------------------------
%{\bf Minimal tight-binding model for bilayer boron nitride.}
\section{A minimal tight-binding model for bilayer boron nitride}
%------------------------------------------------------------------------------------------
\label{sec:theory}
Since we have in mind possible applications where both graphene and boron nitride two-dimensional
crystals are brought into close contact with each other,
we will be most interested in the energy bands due
to the hybridization of the $p_z$ orbitals alone, the so called $\pi-$bands.
The $\pi-$bands of BN can then be studied using an empirical tight-binding approach.
The goal is simple: using DFT methods we will parametrize the tight-binding hopping
and onsite energies of boron nitride layers, which can latter be used for
microscopic calculations, such as tunneling properties.

\begin{figure}
 \includegraphics*[width=0.6\columnwidth]{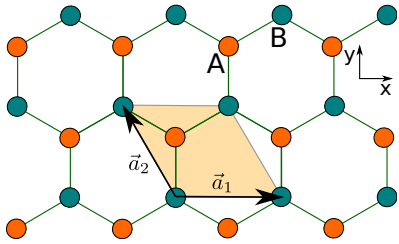}
  \caption{(Color online)
    Illustration of the honeycomb lattice with the $A$ and
    $B$ sublattices, the unit cell, and the primitive vectors $\bm a_1$ and $\bm a_2$.
    It is equivalent to assume that boron or nitrogen is on the $A$ or on the $B$ sublattice.
  }
 \label{fig:unitcell}
\end{figure}

In Fig.~\ref{fig:unitcell} we represent the unit cell of a single layer of boron nitride
with the primitive vectors $\bm a_1$ and $\bm a_2$, both of length $a$.
The minimal tight binding model for a sBN has three parameters
only: the hopping $t$
among nearest neighbor atoms and the onsite energies at the boron and
the nitrogen atoms (in fact,  one of these two onsite energies can be taken as a
reference energy, and one ends up with  two fitting parameters only).
According to this simple model, the Hamiltonian for the electrons
in the $\pi-$bands of sBN can be written as
\begin{equation}
 H =\left[ { \begin{array}{cc}
      E_B & \phi\\
      \phi^* & E_N
     \end{array}}\right]\,,
\label{eq_TBH}
\end{equation}
where $E_B$ is the energy at the boron site, $E_N$ is the energy at the nitrogen site, and
\begin{equation}
 \phi/t =  {1 + e^{ia(-k_x/2 + \sqrt{3} k_y/2)} +  e^{ia(k_x/2 +
\sqrt{3}k_y/2)} }\,.
\label{eq:phi}
\end{equation}
The wave vector in the Brillouin zone is written as $\bm k=(k_x,k_y)$ and
the eigenvalues of Hamiltonian (\ref{eq_TBH}) are given by
\begin{equation}
 E = E_0 \pm \dfrac{1}{2}\sqrt{E_g^2 + 4\vert\phi\vert^2 }\,,
\label{eq:TB}
\end{equation}
where
$E_0=(E_B+E_N)/2
$
is the energy in the middle of the gap and
$E_g = E_B-E_N$
is the energy gap. \cite{Slawinska2010}

Using the above defined tight binding model
one has three parameters that one needs to adjust: the hopping parameter $t$,
the band gap $E_g$, and the middle gap energy $E_0$, which can conveniently be chosen to be zero.
These parameters can be found by adjusting Eq. (\ref{eq:TB}) to
the band structure of sBN
computed from first principles.

For bilayer boron nitride, our minimal model includes,
in addition to the one given above, an interlayer hopping parameter $t'$
between the two atoms that are directly on top of each other.
Hopping between atoms located at larger distances
 can be added to the model, at the expenses of having more fitting
parameters.
If one wants to fit accurately
at the same time the valence and the conduction bands,
adding these extra hoppings is strictly necessary.
However, it is known that conduction bands in insulators are not well described
by DFT calculations and thus we have found not to be necessary to go beyond this minimal model.

\begin{figure}[h]
 \centering
 AA\hspace*{0.3cm}\includegraphics[scale=0.7]{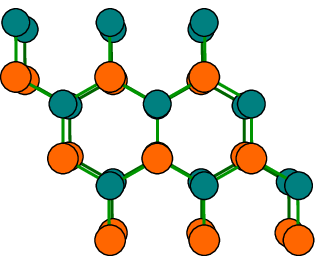}\hspace*{1.3cm}
 AA$'$\hspace*{0.3cm}\includegraphics[scale=0.7]{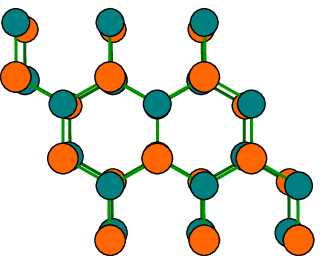}\\
\bigskip
 A$'$B\hspace*{0.3cm}\includegraphics[scale=0.7]{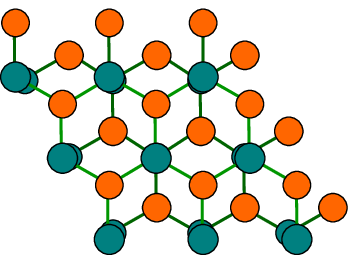}\hspace*{1cm}
 AB$'$\hspace*{0.3cm}\includegraphics[scale=0.7]{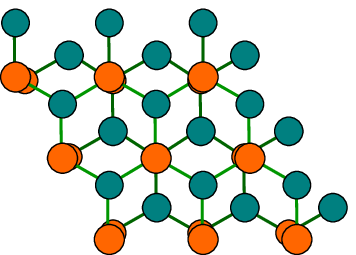}\\
\bigskip
 AB\hspace*{0.3cm}\includegraphics[scale=0.7]{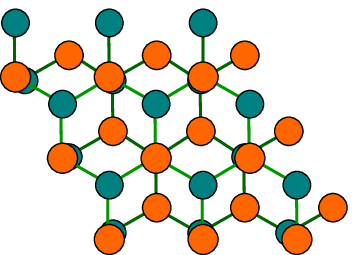} \hspace*{1.5cm}
\includegraphics[bb=0 0 55 26]{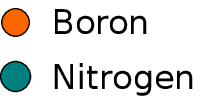} \hspace*{1cm}
 \caption{(Color online) Drawings of the five possible BN bilayers. Some
perspective was allowed to better recognize the type of bilayer.}
 \label{fig:bilayers}
\end{figure}

Five different structures of BN bilayers can exist in principle,
assuming that either the atoms are located exactly on top of each other
or at the center of the hexagons (we are excluding, for example, twisted
bilayers\cite{JLS}); the five structures are
represented in Fig. \ref{fig:bilayers} (in the text that follows we will
be referring to the different structures of BN bilayers by the notation
introduced in that figure, which refer to the possible types of stacking).
The AA and AA$'$ stacking (see Fig. \ref{fig:bilayers})
differ on what type of atoms are superimposed: in the AA case,
atoms of the same type are superimposed, while in
the
AA$'$ case the boron atoms are on top of the nitrogen atoms and vice-versa.
The A$'$B stacking has the nitrogen atoms superimposed on the two layers
and a boron atom at the center of the hexagon, while
AB$'$ stacking has the boron atoms superimposed on the two layers
and a nitrogen atom at the center of the hexagon.
The AB stacking has atoms of different types superimposed on the two layers
and a boron atom at the center of the hexagon in one of the layers, while
on the other layer is a nitrogen atom that is at the center of the hexagon.
The differences among these fives possibilities should be clear from a close inspection
of Fig. \ref{fig:bilayers}.

For boron nitride bilayers one finds that only two different Hamiltonians need to be written,
since they describe all the five possible stacking (see Fig. \ref{fig:bilayers}) of bBN, after the appropriate changes.
Therefore we have:

\begin{equation}
 H_{AA} =\left[ { \begin{array}{cccc}
E_1    &  \phi  &  t'    &  0    \\
\phi^* &  E_2   &  0     &  t'   \\
 t'    &   0    & E_3    & \phi  \\
 0     &   t'   & \phi^* & E_4   \\
     \end{array}}\right]\,,
\label{eq_H_AAP}
\end{equation}
for the first two stacking in Fig. \ref{fig:bilayers} (AA and AA$'$), and
\begin{equation}
 H_{AB} =\left[ { \begin{array}{cccc}
E_1    &  \phi  &  t'    &  0    \\
\phi^* &  E_2   &  0     &  0    \\
 t'    &   0    & E_3    & \phi  \\
 0     &   0    & \phi^* & E_4   \\
     \end{array}}\right]\,,
\label{eq_H_AB}
\end{equation}
for the remaining three cases in the same figure (AB, AB$'$ and A$'$B).
In the last two Hamiltonians
$E_1$, $E_2$, $E_3$ and $E_4$ are the energies at sites 1 to 4, two of which are boron atoms and the other two are nitrogen atoms.
Atoms in the same plane are of different type implying that $E_1\neq E_2$ and $E_3\neq E_4$.
The parameter
$t'$ is the hopping between planes and $\phi$ is the same expression as in Eq. \ref{eq:phi}.
Each of these Hamiltonians give four bands.

For the Hamiltonian (\ref{eq_H_AAP}), cases AA and AA$'$, the four energy bands can be written as:
\begin{equation}
 E = E_0 \pm \dfrac{1}{2}\sqrt{E_g^2 + 4(t'^2 + \vert\phi\vert^2) \pm 8t'\vert\phi\vert  }\,,
\label{eq:AA}
\end{equation}
where $E_0$ and $E_g$ have the same definition as above.
For the Hamiltonian (\ref{eq_H_AB}) we have to consider two different cases.
The first one when $E_1 = E_3$ and $E_2 = E_4$, meaning that there are two atoms  of
the same type (belonging to different layers; cases AB$'$ and A$'$B)
on  top of each other; in this case the energy eigenvalues read:
\begin{equation}
\left\lbrace  \begin{array}{l}
 E = E_0 \pm \dfrac{t'}{2} + \dfrac{1}{2}\sqrt{(E_g\pm t')^2 + 4\vert\phi\vert^2 }\,,\\
 E = E_0 \pm \dfrac{t'}{2} - \dfrac{1}{2}\sqrt{(E_g\pm t')^2 + 4\vert\phi\vert^2 }\,.\end{array} \right.
\label{eq:AB1}
\end{equation}
Note that $E_g = E_B-E_N$ for the case AB$'$ and $E_g = E_N-E_B$ for the case A$'$B.
The second case happens when $E_1 = E_4 = E_B$ and $E_2=E_3=E_N$, where the atoms on top of
each other are of different types; in this case the energy eigenvalues are given by:
\begin{equation}
 E = E_0 \pm \dfrac{1}{2}\sqrt{E_g^2 \pm 2t'\sqrt{t'^2+4\vert\phi\vert^2} +
2t'^2 + 4\vert\phi\vert^2 }\,.
 \label{eq:AB2}
\end{equation}

In  equations (\ref{eq:AA}), (\ref{eq:AB1}), and (\ref{eq:AB2})
 there are four parameters which one needs to adjust: $E_0$, which can be set to zero, $E_g$, $t$, and $t'$.
In the bilayer, $E_g$ is not the energy gap, although it has the same definition as in the
monolayer (see above).
Having described the TB model for bBN, it is necessary to
perform first principle calculations in order to determine the bands of these
systems. Once this is done the TB bands are fit to the {\it ab-initio} results
and the TB parameters defined above are extracted.

The density functional theory calculations were performed with an {\it ab-initio} spin-density functional code ({\sc aimpro}).\cite{Rayson2008}
We have used the GGA in the scheme of Perdew, Burke, and Ernzerhof.\cite{Perdew1996}
Lower states (core states) were accounted for by using the dual-space separable pseudopotentials by Hartwigsen, Goedecker and Hutter.\cite{Hartwigsen1998}
The valence states are expanded over a set of $s-$, $p-$, and $d-$like localized Cartesian-Gaussian Bloch atom-centered functions.
In the framework of DFT calculations, the Brillouin-zone (BZ) was sampled for integrations according to the scheme proposed by Monkhorst-Pack.\cite{Monkhorst1976}
The $\bm{k}$-point sampling was $20\times20\times1$ and both the atoms and the unit cell parameter were relaxed iteratively in order to find the equilibrium positions and size of the primitive cell.
A supercell with hexagonal symmetry was used; the parameter $a$ was varied to find the equilibrium lattice constant, while the $c$ parameter was kept at 60~a.u.

%------------------------------------------------------------------------------------------
\section{DFT results and tight binding fits of the energy spectrum of monolayer
and bilayer boron nitride}
\begin{figure}
 \centering
 \includegraphics*[width=0.8\columnwidth]{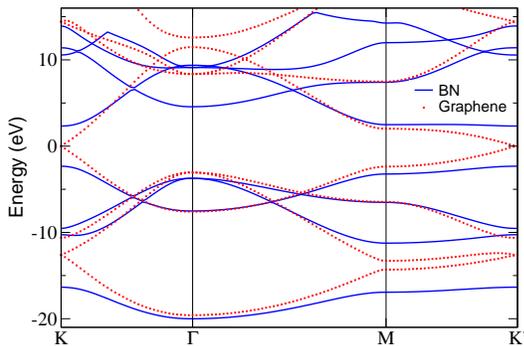}
 \caption{(Color online) Comparison between the band structure of a BN  monolayer and that of graphene.
The solid lines refer to the bands of BN and the dotted curve to the bands of graphene.
As discussed in the text, the main difference between the two band structures occurs
close to the ${\bm K}$ and ${\bm K'}$ points.}
 \label{fig:unitcell_band}
\end{figure}

%------------------------------------------------------------------------------------------
Boron nitride monolayers have
 very different electronic properties from those of graphene, since the broken symmetry of
the $A$ and $B$ sub-lattices necessarily precluded the existence of
 Dirac cones at the corners of the Brillouin zone, creating a large band gap (larger than 5 eV)
in the BN band structure.\cite{Han2008}
Except for this vital difference,
the rest of the electronic band structure of sBN resembles that found for graphene.
The gap is at the $K$-point and not at the $\Gamma$-point (as in usual semiconductors)
and, except for that particular zone in  \textbf{k}-space, the shape of the bands of
BN is similar to that of graphene bands. This is clearly shown
in Fig. \ref{fig:unitcell_band}, where the band structure of the boron nitride single layer is shown
superimposed on the band structure of graphene for comparison.
The lattice parameter we have obtained for BN is $a_{BN}=a=2.51$~\AA{}, in good
 agreement with the value measured experimentally,\cite{Han2008} and for graphene we have found
$a_g=2.46$~\AA{}, giving
a lattice mismatch of the order of 2\% between the lengths of the primitive
vectors of both crystal structures.

\begin{figure}
 \centering
 \includegraphics*[width=0.8\columnwidth]{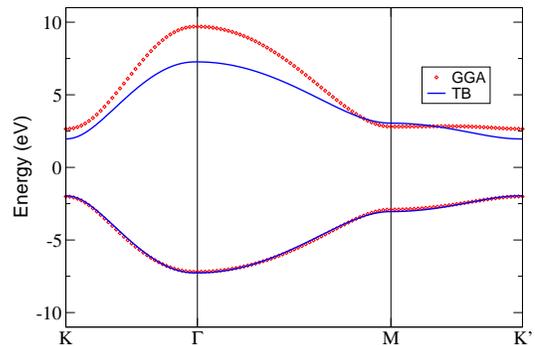}
 \caption{(Color online) DFT band structure of the $\pi-$bands of a BN single layer (solid line)
and of the tight binding fit (dotted line). As
explained in the text, only the valence band was fit.}
 \label{fig:TB_adjust}
\end{figure}

The parameters of Eq. (\ref{eq:TB}) were adjusted to the $\pi$ orbitals of the
valence band only, since DFT is more accurate for occupied states.
The fitting was done to all the band in the  $\bm{K}-\bm{\Gamma}-\bm{M}-\bm{K'}$ line, unlike
the fit done in Ref. [\onlinecite{Slawinska2010}].
In this latter work the bands were fit near the $\bm{K}$-point only.
The curves resulting from fitting Eq. (\ref{eq:TB}) to the {\it ab initio} data
are shown in Fig. \ref{fig:TB_adjust}.
The fitting to the valence band is excellent, but the resulting parameters do not give a perfect fit to
the conduction band, although the TB conduction band does follow the same trend as that
obtained from the {\it ab-initio} calculation, as it should be.
Given that DFT calculations results in conduction and valence $\pi-$bands
that are not particle-hole symmetric (for example,
one has the width of the conduction band greater than that of the valence band),
and that Eq. (\ref{eq:TB}), by construction, preserves that symmetry, the tight binding model
considering only a nearest neighbor hopping parameter will never fit both bands at the
same time.
The tight-binding parameters obtained from fitting the
{\it ab-initio} bands are $E_g=3.92$~eV and $t=2.33$~eV, and the best fit is plotted
in Fig. (\ref{fig:TB_adjust}).
The number we have obtained for $t$ is different from the one ($t$=2.79~eV)
obtained in Ref. \onlinecite{Slawinska2010}  and also from that obtained by the same group in a more recent work ($t$=2.28~eV).\cite{Slawinska2010-1}
In these two works the fitting was done only near de $\bm{K}$-point,
and considering both valence and conduction bands, which is, most likely, the reason for
the discrepancy between ours and the latter result\cite{Slawinska2010-1} by those authors.
We note that in their most recent work\cite{Slawinska2010-1}
those authors have found a value for $t$ closer to the one we are reporting  here.
Our result for $t$ also agrees very well with the one we have obtained for the bBN
(see Table \ref{tab:fit_bilayer});
this is an important consistency check for our calculations.

%------------------------------------------------------------------------------------------
%\subsection{BN bilayer}\label{sec:bilayer}
%{\bf Results for BN bilayer}.
%------------------------------------------------------------------------------------------
We now turn to the study of the electronic spectrum of boron nitride bilayers.
In our DFT calculations
the primitive vector length $a$  was let to relax for all
stackings and its final value was found to be $a=2.51$~\AA{}
for all the cases represented in Fig. \ref{fig:bilayers};
the area of primitive cell size remains essentially unchanged for the five types of stacking
considered here.
The situation is different in what concerns
interlayer distances, as expected.
Table \ref{tab:d} summarizes the results: if different types of atoms are superimposed, the two planes are closer than if the atoms are of the same type, since the chemical bonds
of the type boron-nitrogen are preferred (leading to lower energy states), whereas boron-boron and
nitrogen-nitrogen bonds are not.
The two most stable stackings (AA$'$ and AB) have the same interplane distance.
The interplane distances calculated here are consistent, although somewhat larger,
than the experimentally measured value, which is $\sim 3.3$~\AA{}.\cite{Warner2010,Han2008}

\begin{table}
\centering\caption{Interlayer distances for the bilayer samples, as calculated by DFT.
The central column specifies the two atoms, each of a different layer,
that are considered for the measured distance.}
\begin{tabular}{lcc}
\hline
 Sample & \hspace{1cm}Atoms\hspace{1cm}  & $d$ (\AA) \\
\hline
AA       & B-B   & 3.75 \\
AA       & N-N   & 3.75 \\
AA$'$    & B-N   & 3.57 \\
AB       & B-N   & 3.57 \\
A$'$B    & N-N   & 3.72 \\
AB$'$    & B-B   & 3.60 \\
\hline
\end{tabular}
\label{tab:d}
\end{table}

\begin{figure}
 \centering
 \includegraphics*[width=0.9\columnwidth]{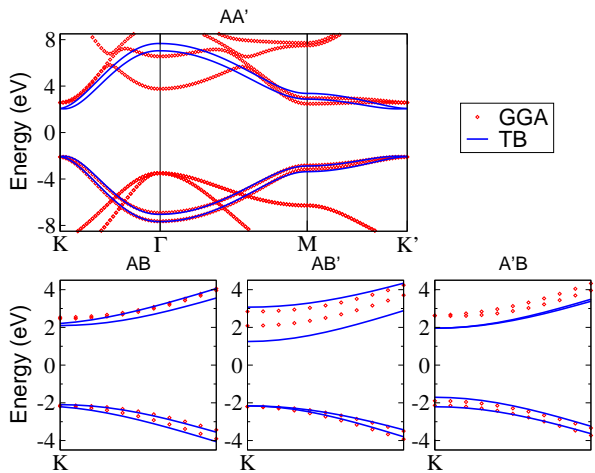}
 \caption{(Color online) Electronic band structure of boron nitride AA$'$ bilayer and details of the electronic band structure near the $\bm{K}$-point for BN bilayers AB, AB$'$ and A$'$B.
The solid line stands for the tight binding fits and the dotted lines for the
{\it ab-inito} band structures. We note that only the valence bands have been fit,
for the reasons given in the text. We cannot expect therefore that the
conduction bands are accurately described by the tight binding energy bands.}
 \label{fig:several}
\end{figure}

Taking the most stable stacking (lowest ground state energy)
as a reference (AB stacking),
the energy difference (per unit cell)
of the corresponding ground states of the
different stackings are given in Table \ref{tab:E}.
The AA stacking is the most energetic, followed by the A$'$B one,
in which  case the nitrogen atoms are on the top of each other.
It is interesting that the most stable stacking
is not the AA$'$ stacking as found for bulk BN,\cite{Terrones2007}
but the AB stacking instead,
as found  for bilayer graphene and  graphite.
These results are consistent with those obtained in Ref. \onlinecite{Marom2010}.
As in that reference, the difference between the ground state
energies of the AA and A$'$B stacking is very small,
meaning, in practice, that both types of stacking can exist; in fact, both have been observed experimentally.\cite{Warner2010}
\begin{table}
\centering\caption{Energy difference, per unit cell,
relatively to the most stable stacking AB and the band gaps
$\Delta_g$ for
 the five bilayer systems, as calculated by DFT.
The third column indicates whether  there is a
degeneracy of the $\pi$-bands at the $\bm{K}$ and $\bm{K'}$ points
in the Brillouin zone
(see Fig. \ref{fig:several}).
In the last column c.b. and v.b. stand for conduction
and valence bands, respectively (see also Fig. \ref{fig:several}).}
\begin{tabular}{lccc}
\hline
          & \hspace{1cm}$E$ (meV)&  \hspace{1cm}$\Delta_g$ (eV) &\hspace{1cm} degeneracy  \\
\hline
AA        & 13.5 & 4.23 &  yes \\
AA$'$     &  0.4 & 4.69 &  no  \\
AB        &  0.0 & 4.60 &  no  \\
A$'$B     & 10.5 & 4.52 &  yes (c.b.)\\
AB$'$     &  3.0 & 4.29 &  yes (v.b.)\\
\hline
\end{tabular}
\label{tab:E}
\end{table}
Table \ref{tab:E} also shows the band gaps for all the bilayers band structure.
The differences among them are small, reflecting the similarity of the band structure of these systems, both qualitatively and quantitatively.
The differences in the band structure of the different types of
bilayers are only noticeable near the $\bm{K}$ and $\bm{K'}$ points, as can be
seen in Fig. \ref{fig:several}.
Figure \ref{fig:several} shows the full band structure of AA$'$ bilayer and the details of the
electronic band structure near the $\bm{K}$-point for the others bilayer stackings,
ploting both the DFT calculations and the tight binding fits.
The tight-binding parameters for each bilayer represent in Fig. \ref{fig:bilayers},
$E_g$, $t$ and $t'$, were adjusted such that
 Eqs. (\ref{eq:AA}), (\ref{eq:AB1}), and (\ref{eq:AB2})
gave the best possible fit to the corresponding band structures as calculated by DFT.
Only the valence band fits are accurate, since they are more reliable on DFT calculations
than the conduction bands (a consequence of the fact that conduction bands have no
electronic density in insulators).
The TB conduction bands, as resulting from the fitting to the valence bands,
are not in quantitatively good agreement with the DFT calculations, but qualitatively do show the same features.
The resulting parameters are shown in table \ref{tab:fit_bilayer}.
It can be seen in Fig. \ref{fig:several} that the fits to the {\it ab-initio}
bands are well described by the tight binding model using the parameters
given in Table \ref{tab:fit_bilayer}.
As in  the case of boron nitride monolayer, the tight binding valence bands describe
quite well the bands calculated by DFT.

Although the three bottom panels of Fig. \ref{fig:several} may not give a clear
impression of the good quality of the fit of the valence bands, given the scales of the figures,
the fits are as good as that seen in the top panel of the same figure, where the
full band structure of the AA$'$ bilayer is shown.
For this reason, the full band structure is plotted only for the AA$'$ stacking,
while for the other cases only the details of the band structure close to the
$\bm{K'}$ are shown.

\begin{table}
\caption{Tight-binding parameters for the
different bilayer stacking.
The negative $E_g$ value for A$'$B results from the exchange between N and B atoms
for that particular stacking.
In the table, $E_g$, $t$ and $t'$ stand for the $E_B-E_N$ ($E_N-E_B$ for A$'$B), the intralayer, and
interlayer hopping parameters, respectively.}
\begin{tabular}{lccc}
\hline
     & \hspace{1cm}$E_g$  (eV)& \hspace{1cm}$t$  (eV)& \hspace{1cm}$t'$ (eV)\\
\hline
AA$'$  &  4.08 & 2.36 & 0.32\\
AB     &  4.19 & 2.37 & 0.60\\
A$'$B  & -3.91 & 2.34 & 0.25\\
AB$'$  &  4.32 & 2.38 & 0.91\\
\hline
\end{tabular}
\label{tab:fit_bilayer}
\end{table}

The values found for the intralayer $t$ hopping parameter are very similar
for the different stackings,
as are the onsite energies at the boron and nitrogen atoms; this is expected
since the intralayer hopping should be
about the same for all stackings as well as for a BN monolayer, and indeed it is.

\begin{figure}
\centering
 \includegraphics[scale=0.5,keepaspectratio=true]{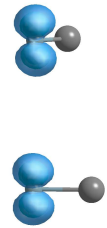}
 % pi_aB.eps: 0x0 pixel, 300dpi, 0.00x0.00 cm, bb=14 14 102 218
 \caption{(Color online) Isosurface for the electron density in the $\pi$-band %at the $\bm{K}$-point
for the BN bilayer A$'$B.  It can be seen that the
electronic density is much higher at the nitrogen atom (smaller spheres) than
 at the boron atoms (larger spheres).}
 \label{fig:pi_aB}
\end{figure}

The $\pi$ valence and conduction bands are degenerated at the $\bm{K}$-point for the AA$'$ bilayer.
For the AB stacking, there is a slight energy difference (at the $\bm{K}$-point)
between the two $\pi-$valence and conduction bands, and therefore no
 degeneracy is observed.
For the A$'$B  bilayer the $\pi-$conduction bands are degenerated at the $\bm{K}$-point,
whereas  for the AB$'$ bilayer the degeneracy occurs for the $\pi-$valence bands.
It can then be  concluded that when a boron atom is on the top of another of the same
species, the $\pi-$valence bands are degenerate at the $\bm{K}$-point, whereas
when a nitrogen atom is on the top of another of the same kind, the reverse happens
and it is the $\pi-$conduction bands that are degenerate. It then comes with no
surprise that for the AA$'$ bilayer both the conduction and valence bands
are degenerate, while for the AA bilayer they are not (these bands are not
shown since they refer to the bilayer with  (by far) the highest ground
state energy when compared to the most stable bilayer).

The electron density is higher around the nitrogen atoms than near the boron atoms (see Fig. \ref{fig:pi_aB}),
which results in a larger distance between the two planes of the bilayer for the A$'$B case.
The $t'$ values shown in table \ref{tab:fit_bilayer} reflect the distance between the layers,
$d=3.72$ \AA{} for A$'$B and $d=3.60$ \AA{} for AB$'$, leading to $t'=0.25$ \AA{} and $t'=0.91$ \AA,
respectively. When different types of atoms sit on the top of each other, the aligned case AA$'$,
with two pairs of atoms superimposed has a smaller $t'$
than the AB case, which is consistent with the lower energy
of the latter.

\vspace{0.5cm}
%------------------------------------------------------------------------------------------
{\bf Final remarks}.
%------------------------------------------------------------------------------------------
The electronic properties of the BN
monolayer and  bilayers were calculated from first principles. Fitting
a minimal tight binding to the DFT band structure the parameters of the
empirical model were determined. The main features
of the {\it ab-initio} bands are well described by our minimal model.
The two most stable bilayer stackings were found to be very close in energy,
suggesting that the system can occur in Nature in both structures.
Since the BN bilayers can be obtained using the
exfoliation method, as in the case graphene bilayers, it is conceivable
that this production method can originate the two different and most stable stacking
BN bilayers.

\end{document}